\documentclass[10pt,journal,compsoc]{IEEEtran}
\usepackage{amsmath,amsfonts}
\usepackage{algorithmic}
\usepackage{algorithm}
\usepackage{array}
\usepackage[caption=false,font=normalsize,labelfont=sf,textfont=sf]{subfig}
\usepackage{textcomp}
\usepackage{stfloats}
\usepackage{url}
\usepackage{verbatim}
\usepackage{graphicx}
\usepackage{cite}      
\usepackage{xcolor}
\usepackage{tabularray}
\usepackage{lscape}
\usepackage{tabularx}
\usepackage{fontawesome5}
\usepackage{booktabs}
\usepackage{multirow}
\usepackage{rotating}
\usepackage{xurl}
\usepackage{orcidlink}
\usepackage{tabularray}

\hypersetup{
  hidelinks
}

\makeatletter

\DeclareTblrTemplate{caption}{paper-class}{%
  \begingroup
    \def\@captype{table}%
    \@makecaption{\fnum@table}{%
      \InsertTblrText{caption}%
    }%
  \endgroup
}

\DeclareTblrTemplate{capcont}{paper-class}{%
  \begingroup
    \def\@captype{table}%
    \@makecaption{\fnum@table}{%
      \InsertTblrText{caption}\space
      \UseTblrTemplate{conthead-text}{default}%
    }%
  \endgroup
}

\SetTblrTemplate{caption}{paper-class}
\SetTblrTemplate{capcont}{paper-class}

\makeatother

\begin{document}

\title{Antipatterns in AI-assisted Qualitative Data Analysis: A Catalog of Temptations and Pitfalls for Software Engineering Researchers}

\author{Rashina Hoda \orcidlink{0000-0001-5147-8096}, Carolyn Seaman \orcidlink{0000-0001-6588-9830}, Victória Gomes \orcidlink{0000-0002-2407-3795}, Rodrigo Spínola \orcidlink{0000-0003-0272-9578}
\thanks{Manuscript received; revised [date].}
\thanks{Rashina Hoda is with Monash University, Melbourne, Australia (e-mail: rashina.hoda@monash.edu). Carolyn Seaman is with University of Maryland Baltimore County, Baltimore, USA (e-mail: cseaman@umbc.edu). Victória Gomes and Rodrigo Spínola are with Virginia Commonwealth University, Richmond, USA (e-mail: oliveiragov@vcu.edu, spinolaro@vcu.edu)}

}

\markboth{Under Review: Hoda et al., 2026}%
{Shell \MakeLowercase{\textit{et al.}}: A Sample Article Using IEEEtran.cls for IEEE Journals}


\maketitle
\begin{abstract}
AI-assisted qualitative data analysis (QDA) offers unprecedented opportunities to streamline software engineering (SE) research, yet uncritical use risks compromising analytical rigor and flooding the field with accelerated production of low-quality research. While tactical best practices will naturally evolve over time, SE researchers currently lack strategic guidance to identify and mitigate methodological risks when attempting AI-assisted QDA. Based on our decades of qualitative SE research expertise and experience combined with an understanding of the emerging landscape of AI-assisted QDA, this paper presents a \textit{catalog of antipatterns in AI-assisted QDA} -- a set of assumptions and practices that initially appear advantageous but ultimately undermine analytical rigor and validity. The antipatterns are grouped into three categories reflecting escalating impact: \textit{Dangerous Drivers}, \textit{Operational Missteps}, and \textit{Analytical Failures}. As more SE researchers attempt AI-assisted QDA, these antipatterns will help them identify and avoid common temptations and pitfalls, while reviewers can be equipped with the vocabulary and criteria to call out problematic and failed practice. Ultimately, this catalog of antipatterns can serve as a stepping stone in our responsible methodological evolution toward principled and meaningful human-AI collaboration in qualitative research.
\end{abstract}


\section{Introduction}
\IEEEPARstart{Q}{ualitative} data analysis (QDA) plays an important role in software engineering (SE) research, enabling rich, contextualized insights into complex and contemporary socio-technical phenomena such as developer use of large language models (LLMs), team dynamics in agile software development, and human-artificial intelligence (AI) collaboration in SE. 

Core principles that guide QDA in SE research include grounding in data, iterative and systematic analysis (e.g., coding and constant comparison) \cite{Charmaz2014, Miles2014}, reflexivity of the researcher in interpreting socio-technical phenomena \cite{Hoda2022, Hoda2024}, and transparency of the analytical process \cite{Seaman1999}. Together, these principles ensure that findings are credible, rigorously derived, and traceable, aligning with established criteria for trustworthiness in qualitative research \cite{LincolnGuba1985}. Rather than merely summarizing or categorizing data, QDA aims to generate explanations, identify patterns, and construct concepts or theory that are deeply grounded in the data \cite{Charmaz2014} and situated within context through interpretive practices \cite{BraunClarke2006}. 

The recent emergence of modern AI (e.g., generative AI and agentic AI) is rapidly transforming how researchers engage with data through QDA techniques such as \textit{thematic analysis} \cite{BraunClarke2006, BraunAndClarke2019}, \textit{socio-technical grounded theory (STGT) for data analysis} \cite{Hoda2022, Hoda2024}, \textit{discourse analysis}~\cite{DiscourseAnalysis}, and \textit{narrative analysis}~\cite{NarrativeAnalysis}. Modern AI tools based on LLMs promise to accelerate the pace of the analytic process, reduce manual effort, and enable the processing of larger datasets than previously possible~\cite{barros2025llmqual,leca2025applications}. In SE, where qualitative methods are often seen to be constrained by time and resource limitations, these capabilities present compelling opportunities and unprecedented temptations to expand the scope and scale of QDA \cite{Hoda2022, LLM4QualInSE}.

However, the integration of AI into QDA also raises significant methodological concerns. When used uncritically, AI can obscure grounding in data, reduce iterative engagement, diminish reflexivity, and weaken transparency, thereby undermining the very principles that ensure the rigor and validity of qualitative research ~\cite{Seaman1999,nowell2017thematic,guba1989fourth, creswellResearchDesign}. Though it offers compelling potential, there is a growing risk that the uncritical use of AI-assisted QDA may enable the production of low-quality qualitative research, faster and at scale, raising concerns about a potential incoming methodological crisis.

This tension is reflected in debates across the wider qualitative research community. At the one end of the spectrum, a group of 419 social science scholars, including pioneering methodologists, have strongly rejected the use of AI in reflexive qualitative research, arguing that interpretive analysis cannot be meaningfully delegated to automated systems without undermining its epistemological foundation such as reflexivity~\cite{WeRejectTheUseOfGenAI}. On the other end of the spectrum, AI adoption is accelerating in research, as evidenced by both the rapid incorporation of AI-based features into established QDA platforms, e.g., MAXQDA~\cite{MAXQDAAI}, NVivo~\cite{NVIVOAI}, ATLAS.ti~\cite{ATLASTI}, and the concurrent emergence of methodological guidelines intended to support AI-assisted qualitative workflows~\cite{ornelas2025llmassistedthematicanalysisopportunities}. This creates a paradoxical situation in which AI-assisted QDA is being simultaneously resisted on principled grounds while also being rapidly promoted in practice.

As socio-technical researchers, it is unsurprising that the SE research community is increasingly intrigued by the prospect of leveraging AI to assist with QDA. Based on emerging work (summarized in the next section) and discussions at SE conferences and workshops, a spectrum seems to be emerging, from researchers and reviewers who are inclined to reject AI-assisted QDA to those who are readily embracing it, with those who are cautiously curious about its use standing in between.

The first two groups represent opposing polarized responses, neither of which is likely to be productive. Embracing AI-assisted QDA without critical reflection risks eroding methodological rigor, while rejecting it altogether risks missing opportunities to address chronic limitations such as speed, scalability, and procedural consistency. Those cautiously curious about AI-assisted QDA seem to be unsure how to proceed due to concerns about invalidating its epistemological foundations and a lack of credible guidance.

Addressing this need for guidance presents an unusual challenge for the SE research community, which has a long history of borrowing, contextualizing, and adapting existing credible guidance on qualitative methodologies from the social sciences to suit our unique socio-technical contexts, e.g., \textit{Case Studies} \cite{runeson2009guidelines}, \textit{Ethnography} \cite{sharp2016role}, \textit{STGT} \cite{Hoda2022}, and \textit{Mixed-Methods Research} \cite{storey2025guiding}. In the case of AI-assisted QDA, there is no existing playbook to follow or adapt from. If anything, prominent social science research experts are largely rejecting AI \cite{WeRejectTheUseOfGenAI}. For a change, as experienced qualitative SE research experts, we may be better positioned to explore the opportunity more thoroughly because of our dual expertise and offer some early guidance to our field.

As a community, we stand on the shore of uncharted waters. Calibrated maps, in the form of tactical guidelines and best practices, will likely emerge in the coming months and years from the attempts, successes, and failures of those undertaking this journey. In the meantime, we cannot rely solely on trial and error. As we begin navigating this emerging space, we need guidance on \textbf{\textit{what can go wrong}} (e.g., epistemological misalignment, loss of reflexivity, bias encoding), \textit{\textbf{why}} such issues arise (based on a deep understanding and hands-on experience of QDA as well as technical expertise in understanding capabilities and limitations of AI), and \textit{\textbf{how to avoid such issues}} when attempting AI-assisted QDA. Such guidance can help move the field from cautious interest toward careful applications. As SE researchers with reasonable understanding of this emerging technology and substantial prior experience in designing, conducting, teaching, and evaluating qualitative research in SE and adapting qualitative research methods for SE ~\cite{Seaman1999, Hoda2022, Hoda2024}, we find ourselves uniquely positioned, and somewhat obliged, to contribute to addressing this early guidance gap. 

In this paper, we aim to help SE researchers and reviewers, regardless of their current stance towards AI, by acknowledging the emerging opportunities and highlighting the temptations, risks, and pitfalls of AI-assisted QDA in SE. To this end, we present a \textbf{catalog of antipatterns in AI-assisted qualitative data analysis}, providing a timely and critical lens for examining where and how methodological issues can emerge. We define antipatterns in this context as \textit{emerging assumptions and practices that initially appear desirable and seemingly reasonable, but ultimately undermine the rigor and validity of qualitative data analysis}. 

We organize eleven antipatterns into three interconnected categories that reflect escalating methodological issues and impact: \textbf{(i) \textit{Dangerous Drivers}}, which capture problematic or unexamined motivations for adopting AI-assisted QDA; \textbf{(ii) \textit{Operational Missteps}}, which represent misconceptions and incorrect or suboptimal uses of AI during the analytical process; and \textbf{(iii) \textit{Analytical Failures}}, which correspond to fundamentally flawed uses of AI-assisted QDA that compromise the methodological integrity and validity of the analysis. 

With this catalog, we aim to provide both researchers and reviewers with a \textbf{structured vocabulary} to communicate and reason about AI-assisted QDA. For researchers, these antipatterns serve as both \textbf{warnings and guidance} to avoid common temptations and pitfalls when attempting AI-assisted QDA. For reviewers, they offer a lens to \textbf{identify and articulate problematic practices} when assessing submitted work. While neither exhaustive nor final, we hope the catalog will contribute to emerging yet critical dialogue and understanding of what constitutes purposeful, meaningful, and valuable QDA in the age of AI~\cite{ornelas2025llmassistedthematicanalysisopportunities,barros2025llmqual}.

The rest of this paper is organized as follows. Section~\ref{sec:background} presents a brief overview of the emerging landscape of AI-assisted QDA as represented in current literature. Section~\ref{sec:catalog} presents the catalog of antipatterns in AI-assisted QDA, including a discussion of their cascading and compounding effects and the scope, application, and limitations of the catalog. Section~\ref{sec:implications} presents recommendations for researchers and reviewers. Finally, Section~\ref{sec:futuredirections} outlines future directions toward responsible methodological evolution in qualitative research in the age of AI.

\begin{table*}[!t]
\centering

\begin{talltblr}[
    caption={Risks associated with AI-assisted qualitative data analysis in software engineering and related areas, classified by the type of evidence reported. {EV:} Empirically validated. {EB:} Experience-based. {UA:} Unvalidated assumptions. {OC:} Opinion-based claims.},
    label={tab:evidence-classification}
]{
    width=\textwidth,
    colspec={
        X[j,m]
        Q[c,m,1.1cm]
        Q[c,m,1.1cm]
        Q[c,m,0.6cm]
        Q[c,m,1.1cm]
    },
    hlines,
    row{4,6,8,10,12}={gray!10},
    hline{1,3,13}={1pt,solid},
    vline{2-5}={solid},
    rows={font=\scriptsize},
    rowsep=0.5pt
}
    \SetCell[r=2]{} \textbf{Risk} & \SetCell[c=4]{c} \textbf{Type of Evidence} \\
    & \textbf{EV} & \textbf{EB} & \textbf{UA} & \textbf{OC} \\
    \textbf{Environmental and planetary sustainability costs:} Training and deploying LLMs require substantial computational resources, which translate into high energy consumption and non-trivial carbon emissions.  & ~\cite{google2024environmental,google2025environmental,shi2025efficient,strubell2019energy} &  &  & \cite{WeRejectTheUseOfGenAI}\\
    \textbf{Bias encoding:} The model may introduce or amplify systematic biases in its outputs.  & ~\cite{barros2025llmqual,Alshaikh2026} & ~\cite{GuidelinesEmpStudies,Ornelas2026,Pizard2026} &  & ~\cite{leca2025applications,BiasInLLM,EthicsAI,ThemeViz,WeRejectTheUseOfGenAI} \\
    \textbf{Loss of depth:} Deep qualitative insight often emerges through in-depth and prolonged engagement with data, careful attention to ambiguity, and sustained effort to reconcile contradictions and tensions within participants’ accounts. Automated shortcuts that accelerate coding or summarization risk compressing this process, thereby reducing opportunities for rich, layered interpretation. & ~\cite{barros2025llmqual} & ~\cite{leca2025applications,Meyer2025Enhancing,Pizard2026} & ~\cite{EthicsAI}  & ~\cite{LLM4QualInSE,GenAiInQualResearch} \\
    \textbf{Loss of reflexivity:} Since reflexivity requires self-awareness, ethical responsibility, and situated judgment, it is an inherently human process that cannot be meaningfully reproduced or automated by LLMs. & ~\cite{barros2025llmqual} & ~\cite{leca2025applications,Ornelas2026} & & ~\cite{bender2021dangers,WeRejectTheUseOfGenAI} \\
    \textbf{Loss of traceability and transparency:} The inherent non-deterministic nature and the frequent version updates of proprietary foundational and frontier LLMs can complicate the ability of future researchers to replicate the process, as the underlying interpretive engine functions as a black box that evolves over time. & ~\cite{leca2025applications} & ~\cite{TowardsEvaluation4ES} & ~\cite{LLM4QualInSE} & \\
    \textbf{Loss of creativity:} LLM-assisted analyses may gravitate toward familiar narratives and commonly observed themes, potentially flattening the creative and exploratory dimensions of qualitative inquiry.  & ~\cite{barros2025llmqual} & & & ~\cite{LLM4SoftEngResearch} \\
    \textbf{Loss of privacy:} Qualitative datasets in software engineering frequently contain sensitive information. When such data are processed using LLMs, particularly via external, cloud-based APIs, there is a risk that sensitive information may be exposed, stored, logged, or reused in ways that are opaque to researchers. &  & ~\cite{rasheed2025largelanguagemodelsserve} & ~\cite{leca2025applications}  &   \\
    \textbf{Loss of human expertise and learning:} When key analytic activities such as coding, memoing, and preliminary interpretation are delegated extensively to LLMs, researchers, particularly novices, may miss critical opportunities to engage deeply with the data and to cultivate these foundational competencies. & & ~\cite{Ornelas2026} & & ~\cite{LLM4QualInSE,bender2021dangers,ThemeViz,LLM4SoftEngResearch} \\
    \textbf{Loss of nuance and diversity:} LLMs generate new codes based on patterns learned from existing datasets, and they tend to favor conventional or statistically dominant interpretations. This creates a feedback loop where researchers inadvertently prioritize existing global concepts over emerging local ones, leading to a loss of nuance and diversity. &  & ~\cite{NguyenTrung2025ChatGPT} & & ~\cite{teixeira2025GenAIQualitative}\\
    \textbf{Loss of emergence:} Emergence refers to the gradual, often unexpected development of concepts, relationships, or theoretical insights that were not specified in advance but arise through iterative engagement with data. This process depends on time, openness, and interaction, conditions that are difficult to preserve when analytic workflows are overly automated or optimized for speed. & & ~\cite{Meyer2025Enhancing} & & ~\cite{GenAiEmpSoftEngineer} \\
\end{talltblr}

\end{table*}

\section{Emerging Landscape of AI-assisted QDA}
\label{sec:background}
The current discussion on AI-assisted QDA across research communities builds on a long history of computational support for QDA, including computer-assisted qualitative data analysis (CAQDAS) tools~\cite{Fielding1998}, topic modeling~\cite{Nikolenko2017}, clustering~\cite{Mitchell2023}, classification~\cite{Crowston2010}, sentiment analysis~\cite{Ingolf2016}, and other machine learning or natural language processing strategies~\cite{Marathe2018}. These perspectives are increasingly consequential because of the kind of analytical work now being delegated to AI. Generative AI and LLMs are increasingly being used in tasks that move closer to the interpretive core of QDA, including \textit{meaning making}, \textit{interpretation}, \textit{code suggestion}, \textit{summarization}, \textit{categorization}, \textit{theme generation}, and \textit{analytic synthesis}~\cite{Shah2026QuaRUM,leca2025applications,LLM4QualInSE}. While traditional software tools, such as older versions of MAXQDA, NVivo, and ATLAS.ti, supported human-led analysis through \textit{structuring}, \textit{cross-referencing}, and some \textit{visualizations}, modern AI systems seem to overtake the interpretive process, \textit{generating codes}, \textit{categories}, \textit{themes}, and even \textit{analytical narratives}. AI is no longer seen only as a low-level aid for organizing qualitative data; it now can shape what is noticed, how patterns are described, which interpretations appear plausible, and how qualitative meaning is constructed.

Table~\ref{tab:evidence-classification} summarizes the risks captured by the emerging body of knowledge in this area, supported by varying levels of evidence. In the social sciences, this shift has led to strong methodological objections, with 419 experienced qualitative researchers from 32 countries, including prominent scholars and methodologists, completely rejecting the use of generative AI for reflexive qualitative research~\cite{WeRejectTheUseOfGenAI}. Their critique is grounded in the argument that QDA depends on \textit{human meaning-making}, \textit{subjectivity}, \textit{situated interpretation}, and \textit{reflexive accountability}. From this perspective, the central problem is not only that AI may produce inaccurate, biased, or superficial outputs, but that it cannot participate in the epistemic work that gives QDA its methodological character. Delegating interpretive tasks to AI may alter the very conditions under which qualitative knowledge is produced, not to mention the overarching issues with accuracy and hallucinations.

Others adopt a less prohibitive, but still cautious, position. Rather than rejecting AI assistance altogether, studies such as Nguyen-Trung~\cite{NguyenTrung2025ChatGPT}, Meyer~\cite{Meyer2025Enhancing} and Teixeira et al.~\cite{teixeira2025GenAIQualitative} examine how AI might support selected analytic activities while preserving human analytical authority. Across this work, AI is commonly framed as a possible aid for \textit{familiarization}, \textit{preliminary coding}, \textit{codebook development}, \textit{summarization}, or \textit{theme refinement}. However, this framing is repeatedly accompanied by concerns that AI may narrow the analytic space, privilege conventional interpretations, reduce engagement with ambiguity, or create a false sense of analytic completeness. \textbf{The central issue, therefore, is whether AI use preserves the depth, emergence, reflexivity, and situated judgment that QDA demands.}

These risks are also being examined within SE contexts ~\cite{LLM4QualInSE,ADESEYE2025,GarciaQuevedo2026,Ngo2026}, driven by practical and methodological tensions. SE researchers often work with large volumes of socio-technical data in textual formats, e.g., interview transcripts, commit messages, developer communication traces, and more recently, AI-generated texts and human-AI conversations \cite{Hoda2024}. These data sources are central to understanding human, organizational, and collaborative aspects of software development, but they also require substantial interpretive effort. Because AI is seen to handle multiple data sources and modalities (e.g., text, audio, video) at scale and speed, modern AI-assisted QDA is emerging as a promising but methodologically delicate concern within SE research.
 
In one of the early works in this area, Bano et al.~\cite{LLM4QualInSE} discuss both the opportunities created by LLMs in qualitative research and the methodological challenges they introduce. They emphasize that QDA depends on contextual understanding, interpretive judgment, and researcher reflexivity, which cannot be directly delegated to automated systems. Leça et al.~\cite{leca2025applications} confirm that LLMs have often been explored for QDA with the expectation that they can reduce manual effort and support the analysis of larger qualitative datasets. However, they also identify limitations related to variability, transparency, privacy, and reproducibility. These concerns highlight the tradeoffs inherent in AI-assisted QDA.

Broader SE work on the use of LLMs in empirical studies also provides some relevant guidance and emphasizes the need to report model versions, prompts, parameters, and usage settings when LLMs are used in empirical research \cite{TowardsEvaluation4ES, GuidelinesEmpStudies}. Although not focused on qualitative analysis, this guidance applies in part to QDA outputs that are also sensitive to prompt wording, model updates, sampling decisions, and non-deterministic behavior.

These risks can be understood as recurring patterns rather than as isolated limitations, i.e., as \textbf{\textit{antipatterns}} -- a diagnostic taxonomy previously used to guide SE researchers on ``\textit{what not to do}'' and ``\textit{what to avoid}'' \cite{Hoda2022, storey2025guiding}.

\section{A Catalog of Antipatterns}
\label{sec:catalog}

\begin{table*}[p]
    \scriptsize
    \caption{A catalog of antipatterns in AI-assisted qualitative data analysis.} 
    \begin{tabularx}{\textwidth}{>{\bfseries}>{\raggedright\arraybackslash}p{0.005\linewidth} >{\raggedright\arraybackslash}p{0.17\linewidth} >{\raggedright\arraybackslash}p{0.17\linewidth} >{\raggedright\arraybackslash}p{0.15\linewidth} >{\raggedright\arraybackslash}p{0.20\linewidth} >{\raggedright\arraybackslash}p{0.18\linewidth}}\label{tab:antipatterns} \\
    \toprule
    \textbf{} & \textbf{Antipattern} & \textbf{Example} & \textbf{Why Tempting} & \textbf{Associated Risks} & \textbf{Possible Mitigation Strategies}\\
    \midrule
    
    \multirow{4}{*}[-2ex]{\rotatebox{90}{DANGEROUS DRIVERS}} 
    & \faHammer~\textbf{Because We Can}: Using AI-assisted QDA because AI is able to process qualitative data and is readily available to human researchers, without due consideration of fit, relevance, risks, and limitations. & A paper reports the use of an AI tool for open coding, describing the tool used, the prompts, the interaction, and the results, without giving a justification for why AI usage was considered and how the AI tool was selected. & SE researchers are technologists at heart, and we are hard wired to explore the limits of new technology. The coolness of what AI can do, along with apparent gains in speed and scale seem sufficient. & \textbf{Loss of privacy, human expertise, learning, reflexivity} and \textbf{environmental and planetary sustainability costs}, along with threats to research validity and missed opportunities for deeper insight, make it imperative to weigh AI’s apparent benefits against its real but less visible costs.
    
    & Researchers should perform and share a thoughtful cost-benefit analysis to ensure AI use is meaningfully justified, rather than assuming speed and scale are sufficient rationales. \\
    \cmidrule{2-6}
    
    & \faStopwatch~\textbf{The Productivity Spiral}: A vicious cycle between increased scale and speed of AI-assisted QDA on the one hand, and increased community expectations of greater scale and speed from qualitative studies on the other.
  & The Anthropic study ``\textit{What 81,000 people want from AI}'' illustrates how large-scale AI-assisted studies can become a visible marker of research ambition and prestige. & Researchers may be subject to pressure to use AI to produce more to keep up with peers and expectations from institutions, funding agencies, and publication venues. & Prioritizing productivity over sustained analytical engagement can lead to \textbf{loss of depth}, \textbf{nuance, diversity}, \textbf{emergence}, \textbf{reflexivity}, \textbf{human expertise}, \textbf{learning}, \textbf{traceability, and transparency}, while normalizing unrealistic expectations.
    
    &  Researchers should use qualitative values (depth, richness, reflexivity) to select and justify their QDA approach, rather than be driven by quantitative markers such as scale and speed.  \\

    \cmidrule{2-6}
    
    & \faCrown~\textbf{Human as Gold Standard}: Assuming human data analysis is an appropriate comparison or benchmark for AI-assisted QDA without considering the expertise and experience of the human analyst(s). & A well-designed empirical comparison finds an AI and a human coder assign similar codes to 90\% of the data and concludes that AI coding is acceptable without considering the human analyst's expertise, experience, and QDA quality. & Manual coding is time-consuming and can be tedious. Getting what seem to be the same results using AI can be seen to justify AI use, saving substantial time. & Benchmarking against  human analyst(s) may be insufficient and misses an opportunity to do better, leading to \textbf{bias encoding}, \textbf{loss of depth}, \textbf{nuance, diversity}, and \textbf{reflexivity}.
    
    & Other avenues for evaluating AI-generated QDA results should be explored, based on known quality dimensions for qualitative research.\\
    
    \midrule
    
    \multirow{5}{*}[-3ex]{\rotatebox{90}{OPERATIONAL MISSTEPS}} 
    & \faTags~\textbf{QDA as Categorization}: Using AI to group data into "buckets" based on internal quantification (e.g., most popular themes), training data, and existing knowledge (instead of staying true to the research data). Limiting QDA to categorization. & Coarse-grained, quantitatively-led categorization missing qualitative richness, e.g., "Types of test debt  identified: missing test cases (mentioned by 60\% participants) and broken test cases (10\%)''.  & Categorization is a low hanging fruit which LLMs seem to be good at (e.g., content analysis). In low risk, non-research contexts (e.g., market analysis), it may be all that is needed. & Reducing QDA to surface-level categorization can drive \textbf{loss of depth}, \textbf{nuance and diversity}; forcing data into fixed buckets can reinforce \textbf{bias encoding} and cause \textbf{loss of emergence} and \textbf{loss of creativity}.
    
    & \textit{Deductive} QDA does not end at categorization; categories should be analyzed for richness and relationships. In \textit{inductive} QDA, AI can assist with individual and incremental steps (e.g., open coding and constant comparison), rather than categorizing across the dataset. \\

    \cmidrule{2-6}
    
    & \faArrowDown~\textbf{Epistemological Slide}: A shift away from the constructivist foundations of qualitative research toward positivism. & AI assumed by some to reflect the `objective truth' and thus being employed as a `superior' analyst to subjective humans. Presentation of findings marked by quantification of qualitative data. & Positioning AI as `objective' and free from human intervention can be used to justify its use in QDA. & Reducing the role of the human in constructing knowledge and over-reliance on AI can lead to \textbf{loss of nuance, diversity, depth, emergence, reflexivity} and \textbf{creativity} and risks \textbf{bias encoding}.
    & Reviewers should watch for and question overly positivist research approaches (quant heavy, claiming to be objective, looking to verify `the truth') when a constructivist approach is more meaningful.  \\
    
    \cmidrule{2-6}
    
    & \faBox~\textbf{The Black Box Trap}: Treating AI as an authoritative, opaque analyst, where generated codes, groupings, and interpretations are accepted at face value without fully understanding either QDA principles or the AI-based mechanisms being used. & A researcher prompts a general purpose LLM to “identify themes” from interview transcripts. The researcher cannot explain the generated themes, how they relate to the data, or whether alternative interpretations exist. & For researchers unfamiliar with QDA, AI can seem like a shortcut around training in methodological principles and steps. Likewise, researchers unfamiliar with AI tools may use them bluntly, without understanding the risks or how to use them well. & Treating opaque AI outputs as authoritative and removing researchers from the analytic process can lead to \textbf{bias encoding} and \textbf{loss of traceability, transparency}, \textbf{reflexivity}, \textbf{human expertise and learning}  while weakening the justification of coding decisions and the grounding of findings in the data. 
    
    & Ensure that researchers have at least a foundational understanding of both QDA and AI tools before integrating them. Require explicit documentation of prompts, intermediate steps, and researcher decisions \cite{TowardsEvaluation4ES}, not merely explanations offered by AI.   \\

    \cmidrule{2-6}
    & \faRocket~\textbf{Too much Trust}: Overreaching analysis caused by hallucinations, generalizations, and over interpretations. & A strong inverse relationship between developer experience and use of AI tools for code generation is found in data analyzed using AI. This is based on stereotypes represented in the underlying general-purpose model and is not present in the research data. & Overreaching analysis can sound grand and impressive, and might be couched in a way that implies more confidence than is warranted.  & Without careful review, AI hallucinations, generalizations, and overinterpretations can misdirect decisions and research agendas, reinforcing \textbf{bias encoding}, \textbf{loss of traceability and transparency}, \textbf{loss of nuance and diversity}, and \textbf{loss of reflexivity}.
    
    & Findings generated using AI should be vetted by humans familiar with the underlying data and with the risks of AI-based analysis. \\

\end{tabularx}
\end{table*}

\begin{table*}[ht]
    \ContinuedFloat
    \scriptsize
    \caption{A catalog of antipatterns in AI-assisted qualitative data analysis (Continued).} 
    \begin{tabularx}{\textwidth}{>{\bfseries}>{\raggedright\arraybackslash}p{0.005\linewidth} >{\raggedright\arraybackslash}p{0.17\linewidth} >{\raggedright\arraybackslash}p{0.17\linewidth} >{\raggedright\arraybackslash}p{0.15\linewidth} >{\raggedright\arraybackslash}p{0.20\linewidth} >{\raggedright\arraybackslash}p{0.18\linewidth}}\label{tab:antipatterns} \\
    \toprule
    \textbf{} & \textbf{Antipattern} & \textbf{Example} & \textbf{Why Tempting} & \textbf{Associated Risks} & \textbf{Possible Mitigation Strategies}\\
    \midrule        
    
    & \faLightbulb~\textbf{AI Output as Insight}: AI-produced QDA findings can sound insightful while remaining shallow, leading researchers and readers to mistake superficial output for meaningful analysis & A researcher uses AI to generate themes from developer interview data about cloud vs on-premises servers. The polished output sounds insightful, but the findings are generic and weakly grounded in data. & AI tools produce fluent, well-structured outputs that resemble legitimate qualitative findings, giving a strong illusion of depth, rigor, and expertise. & Mistaking persuasive AI output for genuine insight creates \textbf{loss of depth} and \textbf{loss of nuance and diversity}; accepting superficial findings without scrutiny also drives \textbf{loss of reflexivity} and \textbf{loss of traceability and transparency}.
    
    & AI-produced results should always be critiqued and guided by human researchers, pushing for depth and nuanced understanding of the underlying data \cite{LLM4QualInSE}.  \\

    \midrule
    
    \multirow{3}{*}[-1ex]{\rotatebox{90}{ANALYTICAL FAILURES}} 
    & \faEyeSlash~\textbf{Low Human Engagement}: Researchers lacking enough familiarity with the raw data to be able to spot mistakes, biases, and hallucinations when applying AI to QDA. & A researcher uses AI to analyze interview data collected by others without reading or becoming familiar with much of it, resulting in hallucinated findings they fail to recognize due to limited engagement with the material. & Reading data is time consuming, and often tedious. Fully reading all the data may seem to erase much of the productivity gains of using AI. & Because AI can produce plausible but incorrect results, detecting errors requires close familiarity with the data. Limited engagement with raw data can therefore reinforce \textbf{bias encoding} and drive \textbf{loss of depth}, \textbf{loss of reflexivity}, and \textbf{loss of human expertise and learning}.
    
    & Read the data. \\ 
    \cmidrule{2-6}
    
    & \faFish~\textbf{Fishing for Results}: Exploiting AI's "eagerness to please" (AKA AI sycophancy) by steering it toward confirming assumptions, prior hypotheses, or external theories rather than staying true to the data. & A researcher believes a proposition is true, but it is not supported by the AI-generated findings. When prompted to consider it, the LLM complies by generating a tenuous chain of "evidence" from the data to support the suggested proposition. & QDA techniques are rigorous, in part, to counter the human tendency to look for the evidence we want in the data (confirmation bias). It is tempting to cede responsibility for countering that tendency to AI when it appears to have found what we are looking for. & We know that human bias is real, and so we have built methodological infrastructure to bound it (e.g., member checking, code reviews). Using AI to legitimize preferred conclusions can bypass these safeguards, reinforcing \textbf{bias encoding} and driving \textbf{loss of emergence}, \textbf{loss of reflexivity}, and \textbf{loss of nuance and diversity}.
    
    & Prompt histories should be reported when AI is used for QDA ~\cite{TowardsEvaluation4ES}. Even more important, prompting techniques should be devised and taught that do not allow researchers to transfer their own biases to AI-based analysis. AI can be trained to be aware of the researcher's \textit{assumptions list} \cite{Hoda2024}. \\
    \cmidrule{2-6}
    
    & \faRobot~\textbf{Automated QDA}: Fully automated, end-to-end AI driven QDA with little to no human engagement with the overall process and analysis. & A research team feeds a set of research questions and a large corpus of interview data into an AI-based QDA tool and asks it to generate a publishable paper on the analysis and results.  & This is taking the efficiency and scale of AI use to the max and exhibits the downsides related to all previous antipatterns. & Fully automating QDA leaves AI decisions and biases unchecked, reinforcing \textbf{bias encoding} and causing \textbf{loss of traceability and transparency}; removing human engagement also drives \textbf{loss of reflexivity}, \textbf{loss of human expertise and learning}, \textbf{loss of depth}, and \textbf{loss of emergence}.
    
    & As a community, we need to develop good workflows that effectively involve AI as a partner in QDA, with frequent and meaningful human-AI collaboration, both for learning and for evaluation. \\
    \midrule
\end{tabularx}
\end{table*}
\vspace{0.3cm}
\begin{center}
\colorbox{gray!20}{
    \parbox{24em}
    {We present a \textbf{catalog of antipatterns in AI-assisted QDA} - \textit{emerging assumptions and practices that initially appear desirable and seemingly reasonable, but ultimately undermine the rigor and validity of qualitative data analysis}. Importantly, these antipatterns should not be read as an argument that AI can never be used to assist QDA. Rather, they identify the conditions under which AI assistance becomes methodologically risky and point to ways to mitigate those risks.} 
}
\end{center}
\vspace{0.3cm}

We believe that there is a place in QDA for effective and prudent AI assistance. Best practices and guidelines in this area will necessarily emerge and evolve as researchers experiment with approaches and critique the results. In the meantime, guidance is needed on \textbf{why not} and \textbf{how not} to apply modern AI tools and techniques in QDA to avoid damage from uncritical or rushed applications. Based on our decades of qualitative SE research expertise and experience combined with an understanding of the emerging landscape of AI-assisted QDA (summarized above), we present a catalog of antipatterns in AI-assisted QDA.

To provide structure and clarity, we organize eleven antipatterns into three categories that reflect the different types of issues that can arise in AI-assisted QDA, from the initial motivation for adopting AI (\textbf{\textit{Dangerous Drivers}}), to how it is operationalized in practice (\textbf{\textit{Operational Missteps}}), to how it can seriously threaten the overall integrity of the analytical process (\textbf{\textit{Analytical Failures}}). Table \ref{tab:antipatterns} presents the details of each \textit{antipattern}, including an illustrative \textit{example}, an explanation of \textit{why it is tempting}, \textit{associated risks}, and some \textit{possible mitigation strategies}. In the following subsections, we summarize the antipatterns in each category, including how reviewers can evaluate these antipatterns with commensurate review recommendations.

\subsection{Dangerous Drivers}

The first category of antipatterns is \textbf{\textit{Dangerous Drivers}}, which are rooted in the \textit{motivation} for adopting AI-assisted QDA. These reflect problematic or unexamined reasons for using AI that can bias methodological decisions from the outset. Although some underlying reasons might be legitimate in some cases, the danger is that the reasons and motivations are usually implicit, thus unexamined, and assumed to be beneficial by default. \textbf{One might think of the antipatterns in the \textit{Dangerous Drivers} category as grounds for reviewers to demand better explanations and methodological justifications for using AI-assisted QDA.}

The \textbf{\textit{because we can}} {\scriptsize\faHammer} ~antipattern is represented by a hammer icon to symbolize \textit{Maslow's law of the instrument}, which says `if you have a hammer, everything looks like a nail'. As technologists, we are embedded in a socio-technical context that encourages technological experimentation, so our proverbial `hammer' is whatever the latest technology buzzword is. Trying something new technologically is rarely questioned and often admired for its own sake. Using AI in QDA because we have access to AI is an easy antipattern to adopt without realizing it. 

Equally subtle, because it can arise from the fear of missing out and a desire to "keep up", is the \textbf{\textit{productivity spiral}} {\footnotesize\faStopwatch} ~antipattern. This arises from the pressures that the ever-increasing use of AI can bring, leading to a vicious cycle between increased scale and speed of AI-assisted QDA on the one hand, and increased community expectations of greater scale and speed from qualitative studies on the other. This phenomenon is not limited to QDA. Concerns about the productivity spiral emerged from Anthropic's analysis of data from 81,000 interviews about what people from all domains and walks of life want and fear from AI~\cite{AnthropicStudy}. Ironically, Anthropic's study itself is an extreme example of this antipattern (also evident from the brag of its scale in the title), as it demonstrates an investigative scale that may make sense for its purposes, but cannot and should not be expected from other qualitative studies. Although the Anthropic study is an obvious outlier, the increases in scale and productivity from the use of AI are often more incremental. 

The \textbf{\textit{human as gold standard}} {\footnotesize\faCrown~} antipattern arises when studies compare AI-generated coding results to human-generated results based on the unquestioned assumption that all human coding is equal and of high quality. In practice, there is a marked difference in the quality of coding produced by individuals with varying levels of QDA expertise and experience. Without assessing the quality of the human coding, it is dangerous to assume that it is an acceptable benchmark. Coding done by a novice human analyst or based on a coding scheme that is mediocre at best hardly qualifies as a benchmark. Even coding judged high-quality cannot be considered the only high-quality coding of a particular corpus of data, so its use as a benchmark is equally suspect. Therefore, the quality of the AI-generated coding that is deemed 'just as good' becomes questionable. Although tempting, benchmarking against any human analyst cannot guarantee good quality AI coding if there is \textit{no or low} actual human engagement in the QDA process -- described later as part of more serious \textbf{\textit{Analytical Failures}}.

\subsection{Operational Missteps}
The second category of antipatterns is \textbf{\textit{Operational Missteps}}, which capture incorrect or suboptimal uses of AI during the analysis process, stemming from misconceptions about the nature of qualitative analysis, the role of the researcher, and the capabilities and limitations of AI. Most instances of antipatterns in this category are fixable because they could be embedded in a valid analytical process, but these misuses weaken overall quality. \textbf{One might think of these \textit{Operational Missteps} as grounds for a "major revision" recommendation on a paper under review.}

The \textbf{\textit{QDA as categorization}} {\footnotesize\faTags~} antipattern employs AI to do what AI actually seems to do really well (categorizing data into buckets), but this does not constitute QDA~\cite{barros2025llmqual}. This is problematic based, first, on how AI performs categorization and, second, on the fact that categorization is only one part of QDA. Most generative AI models approach the categorization of qualitative data through an underlying quantitative approach, surfacing most commonly appearing words or phrases (similar to content analysis) or by relying on their vast underlying training datasets and pre-established conceptualizations (through internet searches). This avoids remaining true to the research data, risking misinterpretations and overfitting meanings ~\cite{barros2025llmqual}. Force-fitting data into context windows (e.g., uploading entire interview transcripts into a single query) does not help. Even when a retrieval augmented generation (RAG) approach is applied or a custom LLM is built, risks such as the loss of nuance, creativity, emergence, and reflexivity cannot be denied (Table ~\ref{tab:evidence-classification}).

Furthermore, reducing QDA to a categorization task is analogous to reducing software engineering to writing code. Writing code is but one important activity in SE, supported by many others such as requirements specification, negotiation, design, architecture, testing, deployment, maintenance, and operations. Software can be produced simply by writing code, but the results are suboptimal in most cases. Similarly, categorization is one step within QDA that is supported by many others such as understanding the underlying data, applying reasoning, analytical questioning of the data, creating and refining codes, constant comparison across the dataset, memoing, raising the data through levels of abstraction, reflexive practice, and critical evaluation \cite{Hoda2024}. 

Does this mean AI can never be used for categorization? No. \textit{Simple categorization} in the form of \textit{coarse-grained} grouping of \textit{simple} data -- such as text responses made up of a few words -- done by AI under human supervision may be suitable in cases where both the likelihood and impact of incorrect classification are low. QDA performed within scholarly studies on non-trivial data is often more complex and nuanced and demands a careful human-led approach~\cite{Hoda2024}.

Similarly, the \textbf{\textit{epistemological slide}} {\footnotesize\faArrowDown} characterizes a shift (whether it be in a particular study or over time) towards increasing levels of positivism at the expense of the foundationally important constructivist origins of qualitative research. This antipattern capitalizes on AI's penchant for quantification, typically a hallmark of positivism. Within a single study, this antipattern might manifest as unnecessary quantification or quantitative language. The assumption that AI is an `objective' entity and hence `superior' to subjective humans, when AI bias encoding is well documented (section ~\ref{sec:background}), further pushes QDA along the epistemological slide. Such assumptions and practices do not serve greater understanding of the questions being studied, and often bring into question why a qualitative approach was chosen in the first place. The role of the human in constructing knowledge and leveraging subjective understanding grounded in lived and professional experience is a \textit{feature} of constructivist research, not a \textit{bug}.

The \textbf{\textit{black box trap}} {\footnotesize\faBox~} antipattern is about using AI tools without really understanding how they work or what their limitations are. For example, one-shot queries that lead to `averagification' of results ignore the need to incorporate context into AI-assisted analysis~\cite{barros2025llmqual}. On the flip side, this antipattern also captures the case where AI is used by researchers to compensate for a lack of expertise, training, experience, or understanding of QDA itself, leading to a variety of methodological weaknesses, such as the \textbf{\textit{QDA as categorization}} {\footnotesize\faTags~} antipattern, described above.

Another type of \textit{\textbf{Operational Misstep}} is placing \textbf{\textit{too much trust}} {\footnotesize\faRocket~} in the tool's output, ignoring its tendency towards reductionism and overstatement~\cite{leca2025applications}. The logical extreme of the \textbf{\textit{too much trust}} {\footnotesize\faRocket~} antipattern is the \textbf{\textit{AI output as insight}} {\footnotesize\faLightbulb~} antipattern: LLMs, in particular, are extremely good at masking a lack of deep insight with well-crafted and intelligent-sounding narratives that are hard to ignore. 

\subsection{Analytical Failures}
Our last category of antipatterns is \textbf{\textit{Analytical Failures}}, i.e., \textit{fundamentally flawed} uses of AI-assisted QDA. These involve breakdowns in the methodological integrity of the analysis process itself, rendering the results invalid by any standard of rigorous qualitative research. \textbf{These antipatterns should typically attract a \textit{reject} recommendation on a paper under review.}

The \textbf{\textit{fishing for results}} {\footnotesize\faFish~} antipattern, for example, results from exploiting AI's tendency to please (also known as AI sycophancy), to agree with suggestions from a human analyst, and even generate "evidence" to support such suggestions, even when those suggestions are so biased as to be spurious. In this way, AI tools can be used to amplify human biases (e.g., confirmation bias) rather than avoid them~\cite{barros2025llmqual}. 

This can be exacerbated by the \textbf{\textit{low human engagement with the data}} {\footnotesize\faEyeSlash~} antipattern, which avoids the analyst's obligation to immerse themselves in the data, to read it, and to be familiar enough with the data to be able to spot errors, biases, and hallucinations generated by the AI assistant. 

Finally, the pinnacle of \textit{\textbf{Analytical Failures}} is the \textbf{\textit{automated QDA}} {\footnotesize\faRobot~} antipattern, which results from all other \textit{\textbf{Analytical Failures}} taken to the extreme. \textit{\textbf{Automated QDA}} {\footnotesize\faRobot~} describes an approach to QDA in which the entire process, from data preparation to writing up of the results, is handed over to an AI tool with no human involvement, no human review of intermediate results, and no opportunity for course correction.

\subsection{Compounding and Cascading Antipatterns}
\label{sec:cascadingAPs}
Antipatterns hardly exist in isolation. Rather, they often co-manifest, imposing a cascading and combined effect. For example, \textbf{\textit{because we can}} {\scriptsize\faHammer~} represents a \textit{technology push} or \textit{AI first} attitude and approach to QDA that risks the manifestation of other antipatterns. AI-enabled QDA can be seen as a silver bullet to overcome chronic limitations of manual QDA such as speed and scale, making antipatterns such as the \textbf{\textit{productivity spiral}} {\footnotesize\faStopwatch~} seem justified and logical. Similarly, \textbf{\textit{human as gold standard}} {\footnotesize\faCrown~} can accompany either or both of these \textbf{\textit{Dangerous Drivers}}. For example, once an AI native approach is deployed, it is easy to fall into the trap of trusting a benchmarking exercise against a human analyst(s), without consideration for their level of QDA expertise, to calibrate and accept AI outputs. \textbf{\textit{Dangerous drivers}} such as \textbf{\textit{because we can}} {\scriptsize\faHammer~} can also cause a cascading effect onto other antipatterns, such as \textbf{\textit{too much trust}} {\footnotesize\faRocket~}, an \textbf{\textit{Operational Misstep}}, or \textbf{\textit{low human engagement with the data}} {\footnotesize\faEyeSlash~}, an \textbf{\textit{Analytical Failure}}.

\textbf{\textit{QDA as categorization}} {\footnotesize\faTags~} is likely to trigger the \textbf{\textit{epistemological slide}} {\footnotesize\faArrowDown~}. In \textbf{\textit{QDA as categorization}} {\footnotesize\faTags~}, concepts \textit{most commonly occurring} in the data are typically grouped to create categories. Such an approach shifts the QDA goal from developing rich and layered understanding and conceptualization toward listing coarse-grained categories that abstract away nuance and may draw on pre-existing ideas from its training. Thus, quantification of qualitative data happens quietly in the background through representing averages, percentages, coverage, means, medians, statistical significance, or even inter-coder reliability statistics. Similarly, even if the \textbf{\textit{human as gold standard}} {\footnotesize\faCrown~} is partially addressed by comparing to an expert human analyst, without continuous and significant human engagement, it can lead to \textbf{\textit{low human engagement with the data}} {\footnotesize\faEyeSlash~} or \textbf{\textit{automated QDA}} {\footnotesize\faRobot~}, more serious \textit{\textbf{Analytical Failures}}. \textbf{\textit{Human as gold standard}} \faCrown~ can also lead to an \textbf{\textit{epistemological slide}} {\footnotesize\faArrowDown~} as the human analyst's view is treated as the benchmark or baseline truth (positivism), without leaving room for multiple interpretations (constructivism). Individually and together, these antipatterns move the epistemological dial of inquiry from constructivism toward positivism.

As mentioned earlier, \textbf{\textit{AI output as insight}} {\footnotesize\faLightbulb~} is a logical extreme of the \textbf{\textit{too much trust}} {\footnotesize\faRocket~} antipattern, while the \textbf{\textit{black box trap}} {\footnotesize\faBox~} can help explain why these two antipatterns may manifest. By treating AI as a black box and not applying critical thinking, human analysts can become prone to \textbf{\textit{too much trust}} {\footnotesize\faRocket~} and consequently accepting any and all \textbf{\textit{AI output as insight}} {\footnotesize\faLightbulb~}. These antipatterns are also likely to form a vicious cycle with \textbf{\textit{low human engagement with the data}} {\footnotesize\faEyeSlash~}, as the human analyst over-trusts and over-relies on AI, but is not familiar enough with the data to recognize when AI has gone off the rails.

The examples above suggest that methodological weaknesses introduced early in an AI-assisted QDA workflow may propagate throughout the remainder of the analysis. Decisions that initially appear minor, such as adopting AI without clear justification or accepting its outputs uncritically, can progressively distance researchers from the data, ultimately compromising grounding, reflexivity, transparency, and the overall trustworthiness of qualitative analysis.

\subsection{Scope, Application, and Limitations}
\label{sec:limitations}

While our goal is to provide a structured vocabulary and analytical foundation for reasoning about problematic attempts at AI-assisted QDA, we do not claim that the proposed taxonomy or the relationships described are definitive or exhaustive. Similarly, some of these antipatterns may arise in other AI-assisted research processes, for example in quantitative research. Nevertheless, this catalog is situated within QDA as the antipatterns are drawn from an understanding of the possible violations of qualitative research principles. \\

\noindent \textbf{Applicability beyond software engineering research.} The authors are all SE researchers and thus are qualified to propose guidance for our own discipline. However, many of the presented antipatterns appear to be more widely relevant. For example, the \textbf{\textit{productivity spiral}} {\footnotesize\faStopwatch~} is observed in the results of the Anthropic interview study as a phenomenon cutting across disciplines. Researchers in related disciplines such as human computer interaction, human robot interaction, information systems, AI, etc. may also find these antipatterns relevant.\\

\noindent \textbf{Evolving nature of AI-assisted QDA antipatterns.} The antipatterns identified in this paper were informed by the capabilities, interaction paradigms, and limitations of contemporary AI systems. However, AI technologies are evolving rapidly, and advances in reasoning, context management, transparency, multimodal capabilities, and autonomous workflows are likely to reshape how researchers interact with these systems in the future. Consequently, some antipatterns described in this work may become less relevant, manifest differently, or even disappear as AI systems mature, while entirely new antipatterns may emerge, e.g., with the proliferation of agentic, neurosymbolic, or other forms of AI in the future. Rather than representing a fixed taxonomy, we, therefore, view the antipattern catalog as a conceptual framework that should evolve over time alongside advances in AI technologies and empirical evidence from their use.\\

\noindent \textbf{Compounding relationships between antipatterns.} Finally, the interactions among antipatterns described in section~\ref{sec:cascadingAPs} are conceptual and intended to illustrate plausible ways in which problematic AI-assisted QDA practices may reinforce one another. In practice, some antipatterns may co-occur more frequently than others, different interaction patterns may emerge, and additional cascading effects may exist beyond those discussed in this paper. \\

These aspects also highlight opportunities for future research. We therefore view this paper not as a definitive or exhaustive taxonomy of AI-assisted QDA antipatterns, but as an initial conceptual foundation intended to stimulate empirical investigation, methodological refinement, theoretical debate, and the development of rigorous human--AI collaboration practices for QDA. We expect formal experimentation with emerging AI tools conducted and reported by the SE community in the future to help improve the conceptual depth and breadth of this catalog.

\section{Recommendations}
\label{sec:implications}

The antipatterns presented in this paper highlight not only ways in which AI-assisted QDA can be problematic or fail, but also how such problems and failures can be recognized and mitigated. In this section, we translate these insights into actionable recommendations for both researchers and reviewers. 

\subsection{Recommendations for Researchers}

The antipatterns provide a lens for researchers to anticipate, recognize, and avoid problematic uses of AI-assisted QDA. By reflecting on the motivations, assumptions, and practices that give rise to these antipatterns, researchers can better safeguard the integrity of their analysis and maintain alignment with the epistemic foundations of qualitative research.\\

\noindent \faChevronRight~\textbf{QDA should follow a human-in-the-lead approach}: A \textit{human-in-the-loop} approach to QDA assumes that the process is AI-driven, with a human analyst primarily as a reviewer or quality assurance auditor of `truth' (rooted in positivism). This ignores the role of the human analyst in shaping the analysis  (constructivism), manifesting an \textbf{\textit{epistemological slide}} {\footnotesize\faArrowDown}. Even assuming highly sophisticated AI tools in the future (i.e., zero hallucination, sycophancy, or inaccuracy), a human simply reviewing AI's steps and outputs will have little influence over the human knowledge building required to construct the findings. They can also suffer from possible cognitive fatigue, cognitive offloading, overreliance, erosion of critical thinking, and loss of motivation \cite{gerlich2025ai, dell2023navigating}. This would offset much of the perceived efficiency gains. Even if the human analyst overcomes these practical challenges and is satisfied with both the AI's process and outcomes, this approach still does not leverage what it means for a human to understand the data, derive meaning, enable emergence and reflexivity, and capture nuance and creativity, which are core QDA values. A human-led AI-assisted QDA approach counters these risks, preserves the values of qualitative research, and celebrates the motivation and joy of research work.\\
    
\noindent  \faChevronRight~\textbf{Build expertise in both QDA and AI before applying AI-assisted QDA}: AI-assisted QDA should not be treated as a shortcut to learning either QDA or AI. To lead QDA, researchers need sufficient methodological and technical literacy to question AI outputs, rather than treating them as neutral evidence or finished findings. This does not mean that researchers must master every technical detail of AI systems. Rather, they need enough understanding to recognize that fluent sounding outputs may still be incomplete, biased, or weakly grounded in the data. They also need enough grounding in QDA to notice when AI use reduces interpretation to categorization, obscures alternative readings, or weakens the connection between evidence and analytic claims. Without such critical expertise, researchers are more likely to engage in \textbf{\textit{QDA as categorization}} {\footnotesize\faTags~}, accept \textbf{\textit{AI output as insight}} {\footnotesize\faLightbulb~}, place \textbf{\textit{too much trust}} {\footnotesize\faRocket~} in the system, or fall into the \textbf{\textit{black box trap}} {\footnotesize\faBox~}. Critical AI use therefore depends on researchers’ capacity to ask informed questions about how AI has shaped the analysis, where its outputs require verification, and when its use may be methodologically inappropriate.\\

\noindent  \faChevronRight~\textbf{Do not give in to peer pressure or the fear of missing out}: Methodological issues in AI-assisted QDA often originate before analysis begins, at the level of the motivations for using AI, often driven by peer pressure and the fear of missing out on new research trends and tools (e.g., from seeing other researchers and publications using AI, even if without clear justification or careful application). Antipatterns such as \textbf{\textit{because we can}} {\footnotesize\faHammer~} and \textbf{\textit{productivity spiral}} {\footnotesize\faStopwatch~} reflect unexamined or convenience-driven reasons for adopting AI that can jeopardize the entire research process from the start. Research design should be built thoughtfully and carefully, with each element thoroughly justified and thought through. The design rationales need to be honest and convincing to those outside the research team. This is true, of course, even when AI is not part of the process, but becomes crucial in the presence of AI because these antipatterns can be subtle and easy to miss. They can lead to premature or inappropriate integration of AI into the research workflow, increasing the likelihood of downstream methodological issues.\\

\noindent \faChevronRight~\textbf{Contain issues from cascading and compounding}: Antipatterns rarely occur in isolation. Issues originating at earlier stages of research, such as problematic motivations, can propagate and manifest as \textbf{\textit{Operational Missteps}} or full \textit{\textbf{Analytical Failures}} at later stages of the research process. Researchers should not only address issues at the point where they become visible, but also reflect on upstream decisions and assumptions that may have contributed. This perspective encourages a more holistic view of methodological rigor, where preventing early-stage temptations and missteps can reduce the likelihood of severe downstream failures.\\

\noindent \faChevronRight~\textbf{Recognize the impact beyond a single study (community concerns)}: Not every antipattern may manifest in every study. However, antipatterns manifesting across different studies over time can lead to the degradation and devaluation of qualitative research in the community. A related concern is the loss of qualitative research skills from the community. Qualitative data analysis skills, if increasingly outsourced to AI, will lead to them eroding from the community's methodological repertoire, one study at a time, one researcher at a time. This in turn erodes the ability of qualitative researchers to remain effective arbiters and evaluators of AI-generated results, even if much of the analytical labor is outsourced to AI. 


\subsection{Recommendations for Reviewers}

The increasing use of AI in QDA introduces new challenges for the evaluation of empirical studies, particularly in assessing rigor, transparency, and validity. The antipatterns identified in this paper provide reviewers with a structured lens and vocabulary to identify and articulate methodological concerns specific to AI-assisted QDA. Rather than focusing on \textit{whether} AI was used, the key task for reviewers is to examine the \textit{risks} that may have been introduced in the analytical process as a result of its application. By leveraging the antipattern catalog, reviewers can more systematically identify and name problematic assumptions, missteps, and failures, and provide more precise and actionable feedback on the credibility of AI-assisted QDA.\\

\noindent \faChevronRight~\textbf{Identify red flags of analytical failure}: Certain patterns should be treated as strong indicators of invalid or methodologically unsound analysis. These include fully automated analysis pipelines, absence of a clear chain of evidence, lack of human engagement with raw data, and findings that are not clearly grounded in the data. Such cases correspond to the \textit{\textbf{Analytical Failures}} identified in this paper and are not acceptable as scholarly qualitative research. Their presence substantially weakens the credibility of the study and should warrant rejection.\\

\noindent \faChevronRight~\textbf{Assess the degree of human engagement in QDA}: Reviewers should pay close attention to the role of the human researcher in the analytical process. Minimal engagement with the data or over-reliance on AI outputs may signal antipatterns such as \textbf{\textit{low human engagement with the data}} {\footnotesize\faEyeSlash~} and \textbf{\textit{automated QDA}} {\footnotesize\faRobot~}. Evidence of sustained human involvement, familiarity with the data, and critical interpretation of AI outputs are essential indicators that the analysis has not been inappropriately delegated to the AI system.\\

\noindent \faChevronRight~\textbf{Focus on traceability and accountability}: A central concern in evaluating AI-assisted QDA is whether the analytical process is transparent and traceable. Reviewers should assess whether the study provides sufficient information to understand how AI was used, how outputs were generated, and how they were interpreted in relation to the data. Lack of traceability and transparency should be treated as a significant limitation, as it prevents meaningful evaluation of the validity and credibility of the findings. In particular, missing details about prompts, interactions, and intermediate analytical steps may indicate deeper methodological issues.\\

\noindent \faChevronRight~\textbf{Use the antipatterns as a diagnostic lens}: For reviewers, the antipattern catalog provides a structured vocabulary for identifying and articulating methodological concerns in AI-assisted QDA. Rather than relying on general impressions, reviewers can use specific antipatterns to diagnose where and how an application may be problematic. For example, inadequate documentation of methods may indicate the \textbf{\textit{black box trap}} {\footnotesize\faBox~}, lack of methodological details about how AI-generated codes, findings, or themes were evaluated by human researchers may reflect \textbf{\textit{too much trust}} {\footnotesize\faRocket~}, and absence of explicit links between data and findings may suggest \textbf{\textit{low human engagement with the data}} {\footnotesize\faEyeSlash~}.\\

\noindent \faChevronRight~\textbf{Evaluate holistically}: Since antipatterns can have cascading and compounding effects, reviewers should look out for how the manifestation of one antipattern may hint at another. This will enable reviewers to evaluate the study more holistically and question both apparent and hidden antipatterns.\\

\noindent \faChevronRight~\textbf{Help authors improve}: As the community begins to experiment with the use of AI for QDA, some grace is in order. Reviewers should avoid knee-jerk responses either for or against the use of AI. Instead, give authors guidance and, where possible, ask for a revision. For example, many instances of \textit{\textbf{Dangerous Drivers}} could be addressed by asking authors to more carefully share their research design rationale. Similarly, at least some \textbf{\textit{Operational Missteps}} can be remedied in a revision that better explains, for example, how human and AI collaborated. The community would benefit from good published examples, and reviewers can facilitate that process with a focus on improvement rather than exclusion. On the other hand, bad published examples can do a lot of long-term damage to the field, so reviewers also need to take their role as gatekeepers seriously and thoughtfully.\\

Ultimately, the goal of this work is not to discourage the use of AI in qualitative research, but to make visible the ways in which its misuse can undermine methodological rigor. By framing these risks through a structured catalog of antipatterns, we aim to support more critical, transparent, and accountable use of AI in qualitative data analysis.

\section{Future Directions}
\label{sec:futuredirections}

The antipatterns identified in this paper highlight some critical issues in the way AI is currently, and potentially, integrated into qualitative data analysis. Achieving rigorous and meaningful \textbf{human--AI collaboration in QDA} requires a shift from reacting to problematic practices toward proactive development of frameworks, methods, and evaluation criteria. In this section, we outline key steps towards this goal, emerging from the limitations and risks exposed by the antipatterns catalog.\\

\noindent \textbf{Improve and articulate intentions.} A recurring theme across the identified antipatterns is the presence of weak or unexamined motivations for adopting AI, often driven by convenience, novelty, or perceived efficiency gains (e.g., \textbf{\textit{because we can}} {\footnotesize\faHammer~}, \textbf{\textit{productivity spiral}} {\footnotesize\faStopwatch~}). Future research should focus on articulating \textit{purposeful} uses of AI in QDA. This includes developing conceptual frameworks to articulate when and why AI should be used, how its role aligns with research objectives, how its use supports the core purpose of doing QDA, and how to distinguish between superficial and substantive contributions resulting from the use of AI in qualitative research. Such work is essential to move the field beyond opportunistic adoption toward intentional and methodologically grounded practice.\\

\noindent \textbf{Design principles for human-AI collaboration.} While current uses of AI in QDA often emphasize efficiency and scale, the antipatterns reveal that such gains can come at the cost of analytical depth, contextual richness, and methodological rigor. Future work should articulate what constitutes \textit{meaningful} human-AI collaboration in qualitative data analysis, focusing on interaction models, division of analytical responsibilities, and workflows that preserve the interpretive and reflexive nature of QDA. In particular, there is a need to better understand how AI can support, rather than replace, core analytical processes, avoiding antipatterns such as the \textbf{\textit{black box trap}} {\footnotesize\faBox~} or \textbf{\textit{automated QDA}} {\footnotesize\faRobot~}. Moving in this direction is critical to ensure that  AI augments, rather than undermines, the epistemic foundations of qualitative research.\\

\noindent \textbf{Develop systematic evaluation guidelines.} The integration of AI into QDA introduces new challenges for evaluating the quality and validity of qualitative studies. There is limited guidance on how to assess AI-assisted analysis, leading to inconsistent and likely extreme or superficial evaluation practices. Future work should focus on developing evaluation and reporting guidelines that explicitly account for the role of AI in the analytical process. These criteria should enable reviewers and researchers to systematically identify antipatterns, assess the degree of human engagement, evaluate traceability and grounding in data, and determine whether AI use contributes to or detracts from the overall quality of the study. Establishing such evaluation mechanisms is essential to supporting consistent, transparent, and rigorous evaluation of human-AI collaboration in qualitative research.\\


Together, these directions point toward a broader research agenda aimed at enabling the responsible evolution of QDA in the age of AI. Identifying antipatterns is a critical first step towards the difficult (and exciting) work of articulating principles, guidelines, and exemplars of what constitutes purposeful, meaningful, and valuable use. By embarking on this journey, the research community will be able to better harness the potential of AI while preserving the methodological integrity of qualitative research.

\section*{Acknowledgments}
We express our gratitude to Margaret-Anne Storey, Christoph Treude, and Fabio Palomba for their thoughtful feedback on drafts.

\bibliographystyle{IEEEtran}
\bibliography{main}
\vfill

\end{document}